\documentclass{aa} 
\usepackage{graphicx}
\usepackage{float}
\usepackage{txfonts}
\usepackage{soul}
\usepackage[colorlinks=true, allcolors=blue]{hyperref}
\usepackage{booktabs}
\usepackage{multirow}
\usepackage{natbib}
\bibpunct{(}{)}{;}{a}{}{,}
\usepackage{footnote}
\usepackage{verbatim}
\usepackage{hyperref}     
\usepackage{lastpage}
\usepackage[normalem]{ulem}
\makeatletter
\renewcommand*\aa@pageof{, page \thepage{} of \pageref*{LastPage}}
\makeatother

\hypersetup{
    colorlinks=true,
    linkcolor=blue,
    citecolor=blue
    }

\newcommand{\rsun}{R$_\odot$\xspace}
\newcommand{\hrieuv}{HRI\textsubscript{EUV}\xspace}

\def\arcsec{$^{\prime\prime}$}

\usepackage{xcolor}

\begin{document}

\title{High-resolution observations of small-scale activity in coronal hole plumes}

\titlerunning{}

\author{
        Ziwen Huang\inst{\ref{i:mps}}\fnmsep\thanks{Corresponding author: Ziwen Huang \email{huangz@mps.mpg.de}}
        \and
        L. P. Chitta\inst{\ref{i:mps}}
        \and
         L. Teriaca\inst{\ref{i:mps}}
        \and
        R. Aznar Cuadrado\inst{\ref{i:mps}}
        \and
        H. Peter\inst{\ref{i:mps},\ref{i:kis}}
        \and
        S. K. Solanki\inst{\ref{i:mps}}
        \and
        T. Wiegelmann\inst{\ref{i:mps}}
        \and
        F. Plaschke\inst{\ref{i:tub}}
}

\color{black}

  \institute{
            Max Planck Institute for Solar System Research, Justus-von-Liebig-Weg 3, 37077 G\"ottingen, Germany\label{i:mps}
            \and
            Institut f\"{u}r Sonnenphysik (KIS), Georges-K\"{o}hler-Allee 401A, 79110 Freiburg, Germany\label{i:kis}
            \and
            Institut f\"{u}r Geophysik und extraterrestrische Physik, Technische Universit\"{a}t Braunschweig,
            Mendelssohnstrasse 3, 38106 Braunschweig, Germany\label{i:tub}
}

 \date{Received 08, 09, 2025; accepted 01, 02, 2026}

\abstract
% context heading (optional)
% {} leave it empty if necessary  
{Largely radial ray-like structures, termed plumes, are often observed in coronal holes. Plumes have been proposed to channel magnetohydrodynamic (MHD) waves and the solar wind into the heliosphere. High-speed propagating disturbances (PDs), though well detected in plumes, cannot yet be clearly assigned to MHD waves or to mass flows. Additionally, plume bases as observed in the extreme ultraviolet are riddled with small-scale transients, which could be related to the PDs.}
% aims heading (mandatory)
{We study three plumes within an equatorial coronal hole observed by the EUV High Resolution Imager (\hrieuv) of the Extreme Ultraviolet Imager (EUI) on board Solar Orbiter. The properties of the small-scale brightenings at the plume bases are investigated to interpret their nature and possible relation with PDs in plumes.}
% methods heading (mandatory)
{A 30 Mm $\times$ 30 Mm subregion is selected for each plume base. We process the images with the Difference of Gaussians (DoG) method to highlight the target brightenings, which are further identified with two different approaches. In the 30-min observation, 50 brightenings are visually selected, which also help to set thresholds for automatic detection, where we find 451 brightenings. Their properties, including apparent velocities on the plane of sky (PoS), are analyzed statistically. Potential field extrapolation based on the magnetic field data from the Polarimetric and Helioseismic Imager on board Solar Orbiter is used for correcting the PoS velocity to the real velocity along the magnetic field.}
% results heading (mandatory)
{We observe that the majority of the base brightenings are small-scale (with an area of less than 1.3\,Mm$^2$), short-lived (lasting less than 5 minutes), and slightly elongated at the plume bases. They display intricate movements, with most exhibiting velocities in the PoS of less than 10\,km\,s$^{-1}$. The velocity distribution of the base brightenings selected via visual identification shows a peak at 6--7\,km\,s$^{-1}$ while that of the automatically detected base brightenings peaks at a smaller velocity of less than 5\,km\,s$^{-1}$. Their 3-dimensional velocities are found to be substantially lower than (and difficult to reconcile with) the speeds detected at greater heights in the plumes.}
% conclusions heading (optional), leave it empty if necessary 
{A direct link between base brightenings and PDs remains inconclusive due to the huge difference between their velocities. We propose two possibilities for base brightenings: they may be related to wave-driven Type I spicules or originate from interchange reconnections. Further investigation is required to validate these hypotheses.} 

\titlerunning{EUI observations of small-scale activity in coronal hole plumes}
\authorrunning{Z. Huang et al.}

\maketitle

\section{Introduction}
\label{sec-intro}

Coronal hole plumes (hereafter plumes) generally appear as extended ray-like structures within a coronal hole. Their observations can be traced back to the early 20th century and a great number of studies have been conducted on plumes \citep[see review by][]{poletto_solar_2015}. Plumes expand super-radially into the outer solar atmosphere to over 30 \rsun \citep{woo_extension_1997, deforest_solar_2001}. They are found to be cooler and denser features compared with the ambient inter-plume regions \citep{ahmad_euv_1977, wilhelm_solar_1998,deforest_solar_2001, wilhelm_solar_2006}. Once a plume is formed, it can last for tens of hours \citep{lamy_characterisation_1997} to days \citep{young_temperature_1999} with successors appearing at the same positions in weeks \citep{deforest_solar_2001}. 

Plumes are observed in visible light \citep[e.g.][]{saito_photometry_1956}, in extreme ultraviolet (EUV) \citep[e.g.][]{ahmad_euv_1977,walker_thermal_1993}, and in soft X-rays \citep[e.g.][]{ahmad_x-ray_1978}. Due to the observing geometry (from or close to the ecliptic plane) and generally large sizes of polar coronal holes (low background), polar plumes are easier to be observed projecting well beyond the limb of the Sun, indicating the inhomogeneity of the outer solar atmosphere and providing information on the general magnetic field of the Sun \citep{shimooda_solar_1958}. On the other hand, tracing plumes in low-latitude coronal holes to greater heights is more difficult because of the much higher background from the nearby quiet sun and, often, active region areas. Nevertheless, these plumes are best observed against the disk, in ultraviolet and EUV emissions \citep[e.g.][]{wang_identification_1995, wang_observations_2008}.

Plumes are found to be closely related to the chromospheric network, which is the pattern outlining the convective supergranular cells. \citet{newkirk_coronal_1968} proposed that the rosette structures at the intersection of the networks lies at the root of plumes. Data from the Solar and Heliospheric Observatory \citep[SOHO;][]{domingo_soho_1995} have revealed strong evidence of the spatial coherence of polar plumes and the network \citep{deforest_polar_1997}. \citet{gabriel_structure_2009} proposed that the observations of ‘curtain plumes’ can be modeled by microplumes appearing at the 5\,Mm width boundaries of the supergranular cells. It has been suggested by \citet{wang_coronal_1995} that the magnetic flux concentrates at the edges of network cells reconnect with the small bipoles pushed there by horizontal supergranular convective flows to produce plumes (also see \citep{wang_network_1998} for the schematic).

\begin{figure*}[htbp]
\centering 
\includegraphics[width=1.0\textwidth]{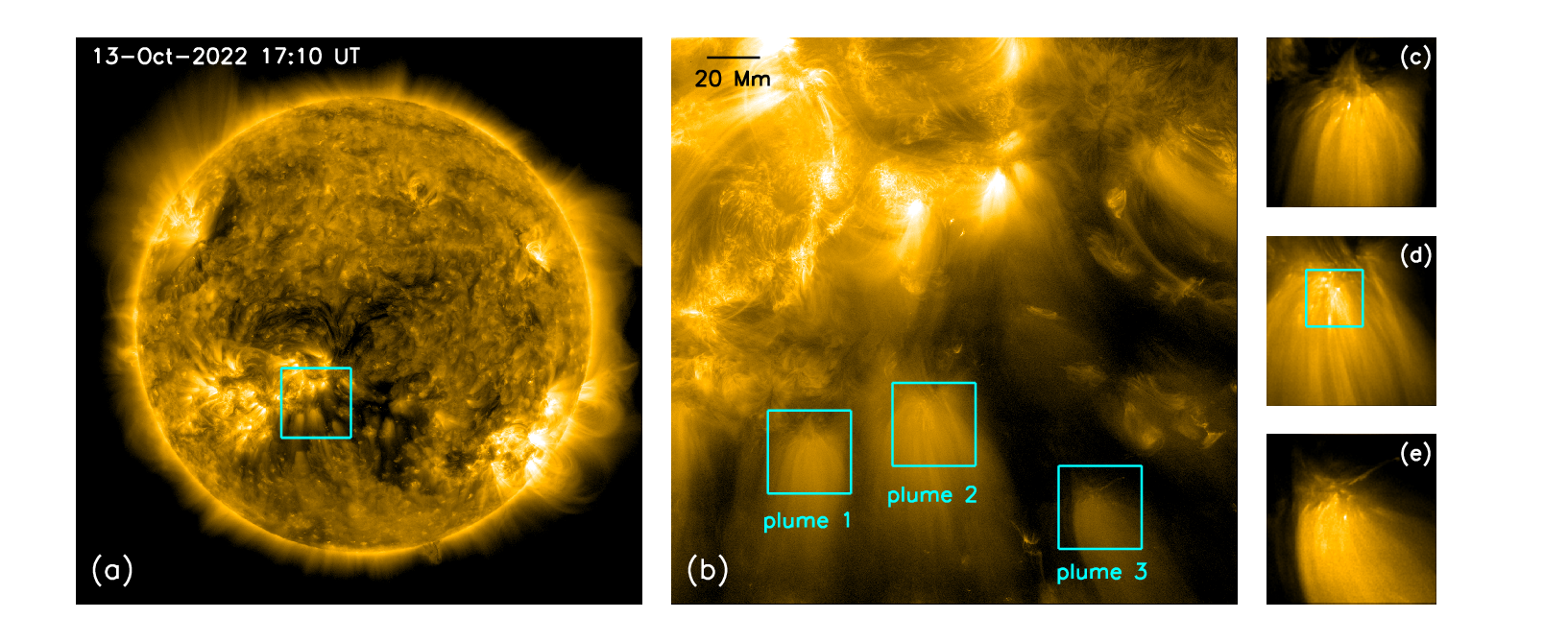}
\caption{Overview of coronal hole observations. (a) Full-disk image of the Sun provided by FSI of EUI. The box indicates the region covered by \hrieuv. (b) Image taken by \hrieuv with three boxes showing the base regions of the three plumes shown in panels (c)-(e). Both (a) and (b) are plotted in logarithmic scale in brightness. (c)-(e) Zoom-in normalized images of the base regions of the three plumes.\\
(The associated movie is available online.)}
\label{obs_fig1}
\end{figure*}

Fast solar wind, which is also believed to originate from coronal holes \citep{krieger_coronal_1973,zirker_coronal_1977,gosling_formation_1999,hassler_solar_1999}, is possibly sustained, at least in part, by coronal hole plumes. Indirect support to this scenario may come from the relation between coronal upflows and network junctions. In fact, \citet{hassler_solar_1999} found a correlation between Doppler shifts in the Ne~{\sc viii} 770 $\AA$ line and chromospheric magnetic network identified in the Si~{\sc ii} $\lambda$ 1533 $\AA$ line. Their work indicates outflows with radial velocity on the order of 5--10\,km\,s$^{-1}$ in the network boundaries and of 10--20\,km\,s$^{-1}$ at the intersections of boundaries. \citet{tu_solar_2005} suggested that the Doppler velocity patches are related to magnetic funnels, which, at least in some cases, could be related to plumes. Despite the above evidences, it remains uncertain whether plumes are a source region for fast solar wind considering the estimation that plumes occupy 10$\%$ of the volume of polar coronal holes, providing only 15$\%$ of the mass \citep{ahmad_euv_1977}.

To help establish the possible contribution of plumes to the solar wind, it is essential to measure the speed of plume outflows at different heights. Numerous studies have been conducted over the years \citep[e.g.,][]{rifai_habbal_flow_1995,corti_physical_1997,wilhelm_solar_1998-1}, with different measurement methods and have yielded widely-spread conclusions. For instance, off-limb plumes are investigated with the Doppler dimming method \citep{noci_solar_1987}. \citet{gabriel_contribution_2003,gabriel_solar_2005} proposed a velocity of beyond 60\,km\,s$^{-1}$ in plumes from 1.05 to 1.4 \rsun, exceeding that in the interplume regions. However, \citet{teriaca_nascent_2003} found that static plumes in the outflowing interplume nicely reproduce the observational lines. In the low corona, some studies measuring Doppler shift of lines in SUMER data failed to retrieve any obvious signature of outflows in plumes \citep{hassler_solar_1999,hassler_spectroscopic_2000,wilhelm_source_2000}. In \citet{feng_stereoscopic_2009}, the stereoscopic reconstructions of polar plumes correct the line of sight (LoS) velocity measured spectroscopically in O~{\sc vi} 1032 $\AA$ and 1038 $\AA$ lines to values along the plume directions that are still below 10\,km\,s$^{-1}$. On the other hand, analyzing observations of plumes in multiple spectral lines by the EUV Imaging Spectrometer \citep[EIS;][]{culhane_euv_2007} on board the Hinode satellite \citep{kosugi_hinode_2007}, \citet{fu_measurements_2014} conclude that plume outflows are quasi-steady with a velocity increasing with height and can be an important source of solar wind. A summary of the outflow speeds both obtained there and from the previous literature can also be found in this paper.

Polar jets, observed at both X-ray \citep{shibata_observations_1992} and EUV \citep{alexander_high-resolution_1999} wavelengths, have been found to be the precursors of plumes \citep{raouafi_evidence_2008}. The Atmospheric Imaging Assembly \citep[AIA;][]{lemen_atmospheric_2012} on board Solar Dynamics Observatory \citep[SDO;][]{pesnell_solar_2012} opened a new era in the study of plumes. Instead of being a spatially coherent ‘haze’, plumes were seen to be formed of filamentary fine structures, the so-called ‘plumelets’ \citep{uritsky_plumelets_2021}. In addition, numerous small jets (so-called jetlets) and plume transient bright points (PTBPs) have been reported to occur in plumes \citep{raouafi_role_2014,pant_dynamics_2015,panesar_iris_2018,kumar_quasi-periodic_2022,chitta_picojets_2023}. \citet{raouafi_role_2014} also proposed that these small-scale energy releases may sustain the evolution of plumes in the long term.

Propagating disturbances (PDs) are observed in plumes as quasi-periodic temporal fluctuations of brightness. They are found to move outwards along the plume filamentary structures with speeds varying from a few 10\,km\,s$^{-1}$ to over 200\,km\,s$^{-1}$. However, their physical nature is not fully understood yet. On the one hand, PDs are thought to be caused by mass flows \citep{mcintosh_stereo_2010,tian_observation_2011}. On the other hand, in many studies they are interpreted to be the signatures of magnetosonic waves \citep{deforest_observation_1998,ofman_slow_1999,banerjee_mhd_2016,banerjee_magnetohydrodynamic_2021}. However, using multi-channel observations, \citet{pucci_birth_2014} found that the PDs have higher speeds at lower temperatures, which is inconsistent with the wave scenario. More recently, \citet{kumar_quasi-periodic_2022} suggested the coexistence of strong flows and waves in plumes. 

Several recent studies have focused on the dynamic processes at the bases of plumes and their connection to jetlets and PDs along plumelets. In particular, SDO/AIA and IRIS observations have revealed quasiperiodic energy releases with characteristic periods of 3--5 minutes in coronal-hole plumes \citep{kumar_quasi-periodic_2022}. They identified recurrent brightenings at plume bases and the associated jetlets moving along plumelets. They also examined the magnetic topology, and proposed a mechanism describing how quasiperiodic reconnection drives jetlet formation. In addition, the periodicities of these base reconnection events match those of microstreams and switchbacks observed by PSP, suggesting a potentially common origin for both phenomena \citep{Kumar_new_2023}. Both works also show the chromospheric/transition region counterparts of these jetlets in AIA 304\AA, which appear to be less elongated than their hotter components. These findings are consistent with \citet{Uritsky_Self_2023}, who suggest that these outflows are generated by impulsive interchange reconnection and likely constitute the primary source of mass and energy for the fast solar wind. However, recent MHD simulations have revealed that interchange reconnection is not a requirement for plume related jetlets. In particular, energy conversion at chromospheric and coronal interfaces of unipolar magnetic field threading plumes could play an important role in driving the upflows \citep{Bora_simulation_2025}. 

Nevertheless, various studies have shown that the fine internal structures in plumes, such as brightenings, jetlets and plumelets, might be intrinsically linked and drive the formation and evolution of the entire plume and, thus, possibly contribute to the solar wind. However, due to resolution limitations, previous investigations on brightenings at plume bases have focused primarily on their temporal intensity variations. Unraveling the detailed morphological evolution of base brightenings and their direct link to plume outflows requires observations with high spatial and temporal resolution. Observations by the ESA and NASA Solar orbiter mission \citep{muller_solar_2020}, acquired near one of the closest perihelia at about 0.3\,astronomical\,units (au), provide unprecedented information about the fine-scale structuring and small-scale dynamics in plumes. Given that the nature of small-scale activities at plume bases is not yet conclusively known, we refer to them as ‘base brightenings’, based on our observations. 

In this work, we focus on analyzing the characteristics and movements of these dynamic base brightenings using high-resolution data from the High Resolution Extreme Ultraviolet telescope (\hrieuv), which is part of the Extreme Ultraviolet Imager \citep[EUI;][]{rochus_solar_2020} on board Solar Orbiter. More details about the observations can be found in section \ref{sec-obs}. In section \ref{sec-ana} we explain two different methods of capturing base brightenings and describe all analyzed properties. This is followed by section \ref{sec-res}, where we show the statistical results and compare velocities of both base brightenings and PDs. The correction of plane of sky (PoS) velocities into the effective velocity along field lines is presented in section \ref{sec-vel}. Discussion about the nature of both base brightenings and PDs can be found in section \ref{sec-dis}. Finally, we summarize our findings in section \ref{sec-con}.

\section{Observation}
\label{sec-obs}

To study the base regions of plumes, we select an equatorial coronal hole captured by EUI on 13 October 2022 adjacent to active regions. Three plumes were embedded in the observed coronal hole section, south of NOAA 13105 and west of NOAA 13107. The observations were taken when Solar Orbiter was at a distance of 0.29\,au from the Sun, and at an Earth Ecliptic longitude of about -107\textdegree. With its 17.4\,nm filter, the \hrieuv instrument recorded a 30-min long imaging sequence from 17:00:00 UT to 17:30:00 UT, at a cadence of 5\,s with an exposure time of 2.8\,s. The high temporal resolution is decreased to 20\,s after binning 4 images together to reduce the noise level during data preparation. \hrieuv images have a angular image scale of 0.492\,\arcsec\,pixel$^{-1}$, corresponding to about 100\,km\,pixel$^{-1}$ (spatial resolution of about 200\,km) at 0.29\,au. The same dataset was used by \citet{2025A&A...694A..71C} to investigate small jets in interplume regions.

In Figure~\ref{obs_fig1}~(a), we show an image of the Sun provided by the Full Sun Imager (FSI) of EUI on Solar Orbiter at 17:10:50~UT. FSI observed the full disk of the Sun at a spatial resolution of about 1800 km and a cadence of 10 min. Available observations  cover the whole lifetime of these plumes. Since here we focus on the small dynamic activity in solar plumes, we mainly use high resolution images from \hrieuv. The box in panel (a) indicates the region covered by \hrieuv, shown in panel (b). To enhance the plumes, which, even though brighter than their coronal hole background, are still much fainter than the surrounding active regions and quiet sun areas, we plot the images in panels (a) and (b) in logarithmic scale. The zoom-in images of the base regions of the three plumes are shown in panel (c)-(e). 

As shown in Figure~\ref{obs_fig1}~(a), these three plumes are rooted in an equatorial coronal hole in the southern hemisphere. The positions of these plumes are close to the disk center with a heliocentric separation angle of about 25\textdegree, which allows observations of the base region with relatively low distortion. Meanwhile, at greater heights, the strong magnetic fields of the active region to the north could push the plume magnetic field southward. Thus, the resulting EUV emission from the plume is apparently more parallel to the PoS, which is better suited to study the movements of the fine structures in the plumes. The FSI sequences show that all the plumes have been existing for more than 1 day and will disappear in about 20 hours after this \hrieuv observation period while plume 3 appears to be already in its decay phase.

Comparing the three plumes we find that features labeled 1 and 2 share similar characteristics that both of them are located near the boundary of the active region with north to south orientation. Plume 3 is rooted in the interior section of the coronal hole and is oriented in a north-eastern to south-western direction. Plume 1 and 2 are also brighter than plume 3. Considering that plume 3 is already decaying, this agrees with the suggestion in \citet{pucci_birth_2014} that plumes die as their gas density is reduced. Despite these differences, all the plumes have highly-dynamic base regions with brightenings appearing near their footpoints and moving outwards.

\section{Data Analysis}
\label{sec-ana}
\subsection{Visual identification Method}

\begin{figure*}[htbp]
\centering 
\includegraphics[width=1.0\textwidth]{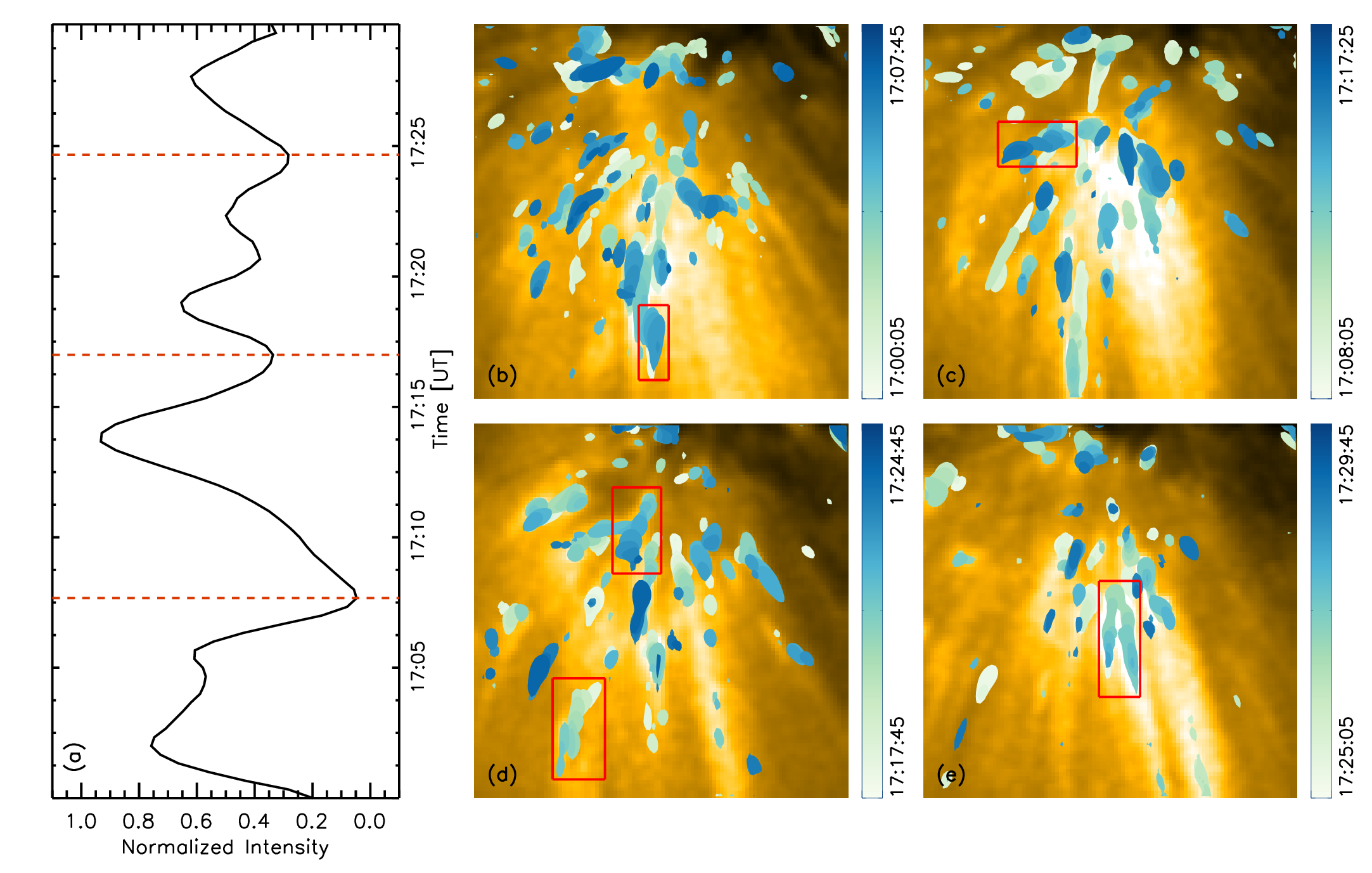}
\caption{Dynamics at the base of plumes. (a) A sample light curve obtained from the base region of plume 2 (see Figure~\ref{obs_fig1}~(d)). The whole time period (30 min) is divided into 4 sections (partitioned at orange dashed lines). (b) - (e) Images of the base region of plume 2 acquired at the middle instance from each temporal section in (a). Brightenings identified using chosen thresholds are marked by the contours (colors indicate time progression; as defined by the color bars). The red boxes highlight several examples of the ‘gradient patterns’, which are formed by contours from successive frames being adjacent to or overlapping one another, thereby illustrating the movement of the brightenings. Note that, even though we have divided the entire observation into four sub-periods, the contour of the brightenings that appears later will inevitably overlap the ones that appear earlier. For example, a discrete contour could be either a brightening that only appears in one frame, or a brightening that does not show a significant movement.}
\label{ana_fig1}
\end{figure*}

To better capture the moving bright features in the base regions of the plumes, we first apply a method based on visualization. Before detailed analysis, we prepare the data by averaging over 4 consecutive images and then applying a running average over $3\times3$ pixels to each resulting image to reduce the noise. We normalize each base region (see Figure~\ref{obs_fig1}~(c)-(e)) to the maximum and minimum intensities of the whole sequence in each sub-region to make the data from the three plumes directly comparable. {Here, we apply the Difference of Gaussian (DoG) method, which is an image processing technique used to enhance edges or features in an image by highlighting areas of rapid intensity change \citep{Lindeberg_scale_1994,lowe_distinctive_2004}. It involves subtracting two Gaussian-blurred versions of the same image. In our study, we use two Gaussian kernels with $\sigma=3$\,pixel and $\sigma=5$\,pixel, respectively. Different thresholds on DoG maps have been tested and we use 0.01 (see Appendix \ref{app_a}) as a reasonable number to cover brightenings as much as possible while avoiding connections of brightenings close to each other. 

Contours representing the target base brightenings at different times are overplotted on the same images to show the trajectory of them. Figure~\ref{ana_fig1} shows plume 2 as an example. To decrease overlap, the entire observational sequence (30\,min) is divided into four distinct time intervals, each covering specific enhancements in the light curve of the base region (see Figure~\ref{ana_fig1}~(a)). Panels~(b)-(e) show the brightening contours obtained with the thresholds of 0.01 on DoG maps in four different time spans. Brightenings appearing at different times are marked with different colors (see color bars in Figure~\ref{ana_fig1}). In these images, the ‘gradient patterns’ (the overlapping contours with changing colors, see examples in red boxes in Figure~\ref{ana_fig1}) indicate the moving structures. Using these ‘gradient patterns’ as a guide to the eye, we check the images to pick up moving brightenings. A few brightenings existing in two time spans are checked throughout their whole lifetimes and still counted as the same ones. With this method a total of 50 base brightenings is selected in the three plumes. 

For each base brightening, a local background, calculated around its position as the minimum value at each pixel along its lifetime, is subtracted. After removing the background, the base brightenings show up better and their boundaries are then re-defined by using a new threshold of 0.75 of their maximum value, after several thresholds being tested (see appendix~\ref{app_a}).

\subsection{Automatic Method}

\begin{figure*}[htbp]
\centering 
\includegraphics[width=1.0\textwidth]{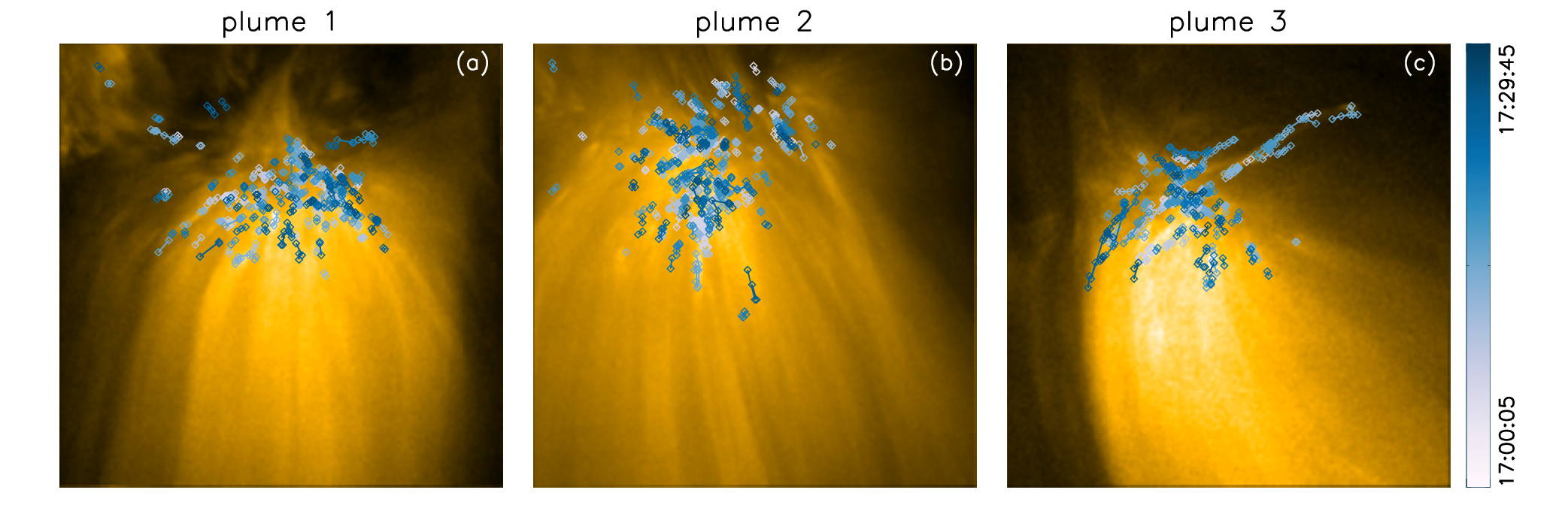}
\caption{Zoom into the plumes bases observed on 13 October 2022 17:10 UT. The intensity-weighted center of each base brightening along the evolution is overplotted on the images. The color represents the time of the first appearance of each base brightening (see the color bar).}
\label{ana_fig2}
\end{figure*}

The visual identification method allows us to select only those base brightenings which are relatively large and bright and experience noticeable shifts in positions. This likely introduces a selection bias. To mitigate this, we also implement an automated selection method to detect brightenings in the three plume bases. Same as for the visual identification method, we apply the DoG method to each averaged image and start with a threshold of 0.01 in the enhanced images. All pixels with a value higher than this threshold are labeled as 'bright pixels'. In each frame, connected bright pixels are considered as a single brightening by doing a 4-neighbor search. Small fragments of less than $2\times2$ pixels are ignored. 

For each detected brightening, we examine the subsequent image frame to track its temporal evolution. If a new brightening larger than $2\times2$ pixels appears within a defined neighborhood of the previous brightening, we interpret both as part of the same evolving structure observed at different times. This neighborhood is defined as a region centered on the peak of the brightening in the previous frame with a radius of 6 pixels. The process is repeated in subsequent frames based on the newly-evolved brightening until the brightening becomes unrecognizable. Considering the pixel size of 100\,km and time cadence of 20\,s, the threshold of 6 pixels corresponds to a velocity of 30\,km\,s$^{-1}$. More details about the threshold selection can be found in appendix~\ref{app_a}. We ignore the brightenings that appear in only one frame. 

A total of 451 base brightenings are picked up with this method. Their boundaries are also re-defined as 0.75 of the maximum value of the base brightenings after the local background being removed. The positions of all base brightenings are shown in Figure~\ref{ana_fig2} where diamonds indicate the positions of the intensity-weighted centers of each base brightening over its temporal evolution with the color representing its start time. The detected base brightenings are also compared with those we picked up by visual identification and 48 out of 50 of those selected manually are also captured using the automatic method. However, fluctuations in brightness over time and the spatial proximity of different base brightenings can result in discrepancies in some events.

\subsection{Properties}

For all the base brightenings selected with both methods, we investigate their properties (area, lifetime, intensity, length/width ratio and velocity) statistically. Based on their boundaries defined on images with background subtracted as described above, we can easily get the area and average intensities, which are defined at the maximum along the evolution of each base brightening. Here we calculate the mean intensity with and without removing the local background. The lifetime is the time difference between the first and last appearance of a base brightening with an accuracy of 20\,s. This lifetime can be considered as an upper limit because it assumes that the base brightening appears at the beginning of the first exposure and disappears at the end of last one. On the other hand, the lifetime of the base brightenings existing already at the start or remaining at the end of the timeseries is underestimated. Considering that only 5\% of the events are influenced and the longest-lived ones have a lifetime shorter than 5\,min, this effect can hardly have a significant impact on the statistical results.

We also study the shape of the base brightenings, which is quantified as the ratio of length to width when the area reaches its maximum along the evolution. This is calculated by fitting the outline of a base brightening with a 2-dimensional Gaussian function, and getting the ratio of the full width at half maximum (FWHM) at both directions, namely, the major and minor axes, as a proxy.

The velocity, representing an apparent PoS motion of the base brightening, is calculated using its initial and final spatial locations. Considering that the size, shape and brightness are changing during its evolution, it is complex to find one representative point for a base brightening. We compare the positions of peak points, the centroid and intensity-weighted centroid and find that statistically they are close enough to have negligible impact. We use the intensity-weighted centroid as the location of the base brightening to estimate the velocity, as it factors in both the extent and intensity.

\section{Results}
\label{sec-res}
\subsection{Visual detection results}

We investigate several properties of the 50 base brightenings selected with the visual identification method described above, including intensity, lifetime, area, shape and velocity. The distributions are shown in Figure~\ref{res_fig1}. The panel (a) indicates that the mean intensities of the base brightenings show different peaks in different plumes. However, as seen in panel (b), they merge closer when the local background is removed. This tells us that although the base brightenings in bright plumes have higher intensities than those in faint plumes, the difference is mainly caused by the overall brightnesses of the plumes.

\begin{figure*}[htbp]
\centering 
\includegraphics[width=1.0\textwidth]{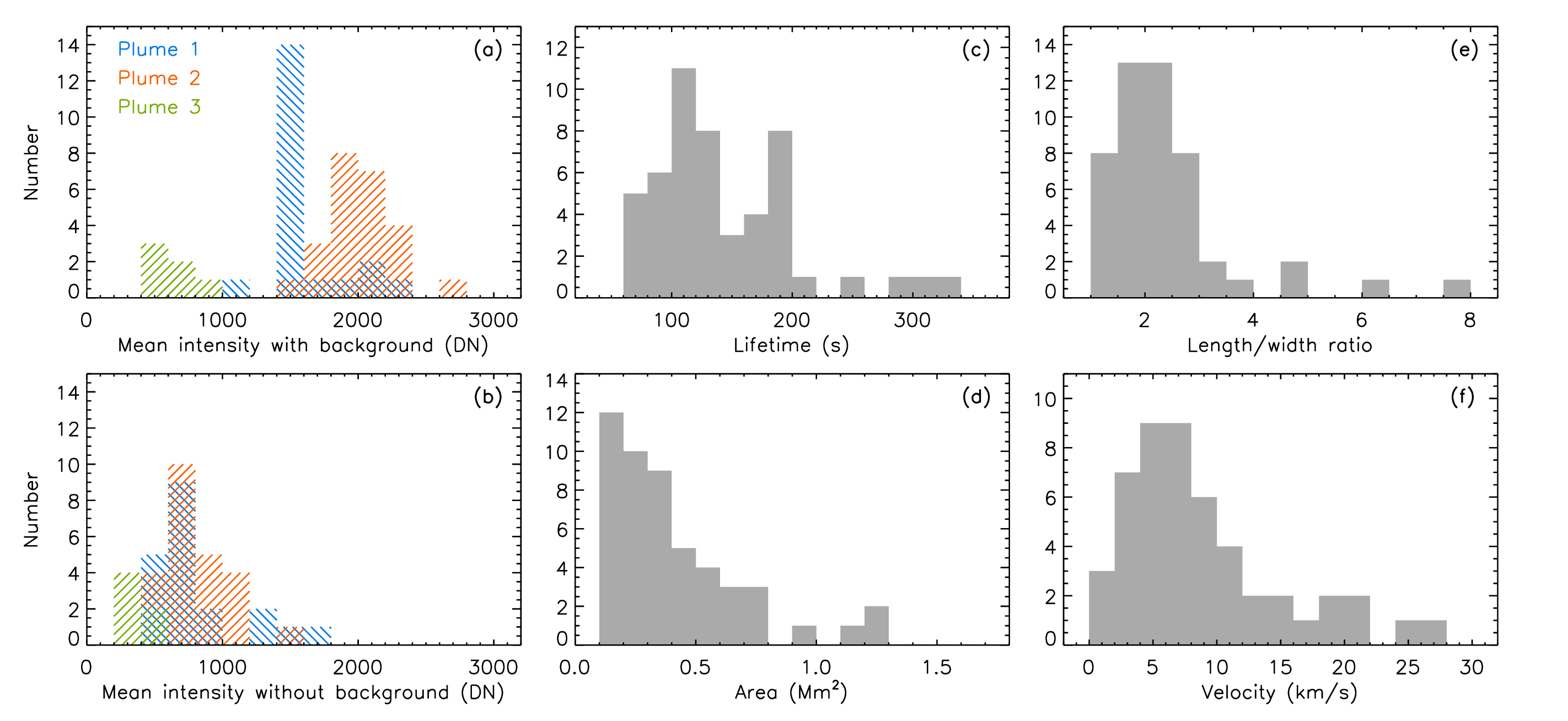}
\caption{Distributions of the properties of the base brightenings selected with the visual identification method (50 base brightenings). (a) shows the mean intensity of the base brightenings without subtracting the local background while (b) shows the mean intensity with the local background removed. (c) - (f) show the lifetime, area, ratio of length to width and velocity. The velocity here is a projection on the PoS and is calculated using the first and last positions of the intensity-weighted center of the base brightening.}
\label{res_fig1}
\end{figure*}

Most of these base brightenings are transient and small-scale, as shown in Figure~\ref{res_fig1}~(c) and (d). They mostly have a lifetime of less than 5 minutes. The short-lived ones have a lifetime of 60\,s. Note that the method we use only allows us to spot brightenings appearing in several images consecutively. The area of the studied base brightenings varies from 0.05 to 1.3 $Mm^2$. These values are based on the threshold used to demarcate the brightenings' boundaries. The distribution of the length to width ratio (Figure~\ref{res_fig1}~(e)) shows a clear peak at around 2 and can go up to 8. This means most base brightenings exhibit a sightly elongated elliptical morphology while a minority exhibit significant elongation, forming a plumelet-like structure at low altitudes.

Figure~\ref{res_fig1}~(f) shows the distribution of the velocity. About $70\%$ of the base brightenings have a velocity component in the PoS of less than 10\,km\,s$^{-1}$. The velocity histogram peaks at around 6\,km\,s$^{-1}$. By closely examining the trajectories of these structures, we find their movements to be more complex than unidirectional following the plume threads. Typically, the base brightenings appear closer to the foot point and move outward, away from the base region. While some of the structures move in alignment with the plume (i.e., from north to south in plumes 1 and 2), some deviate from this direction. Considering a funnel structure of an open magnetic flux tube, in which the magnetic field lines starting from a concentrated region in the photosphere expand nearly horizontally in the chromosphere and transition region, forming a canopy, before they become more vertical again on the way to the corona \citep{Solanki_how_1990, Schrijver_the_2003}. There could also be small-scale loops in the neighborhood in addition to these open field lines. In such a geometry, depending on which field line the base brightening is moving along and at which height, it can show different apparent movements in PoS. Furthermore, both the intensity of the base brightenings and the directions of their motion can change throughout their evolution owing to the variations in photospheric magnetic field on short timescales of less than 5\,min \citep[e.g.,][]{2017A&A...598A..47A,2023ApJ...956L...1C}. 

\subsection{Automatic detection results}

\begin{figure*}[htbp]
\centering 
\includegraphics[width=1.0\textwidth]{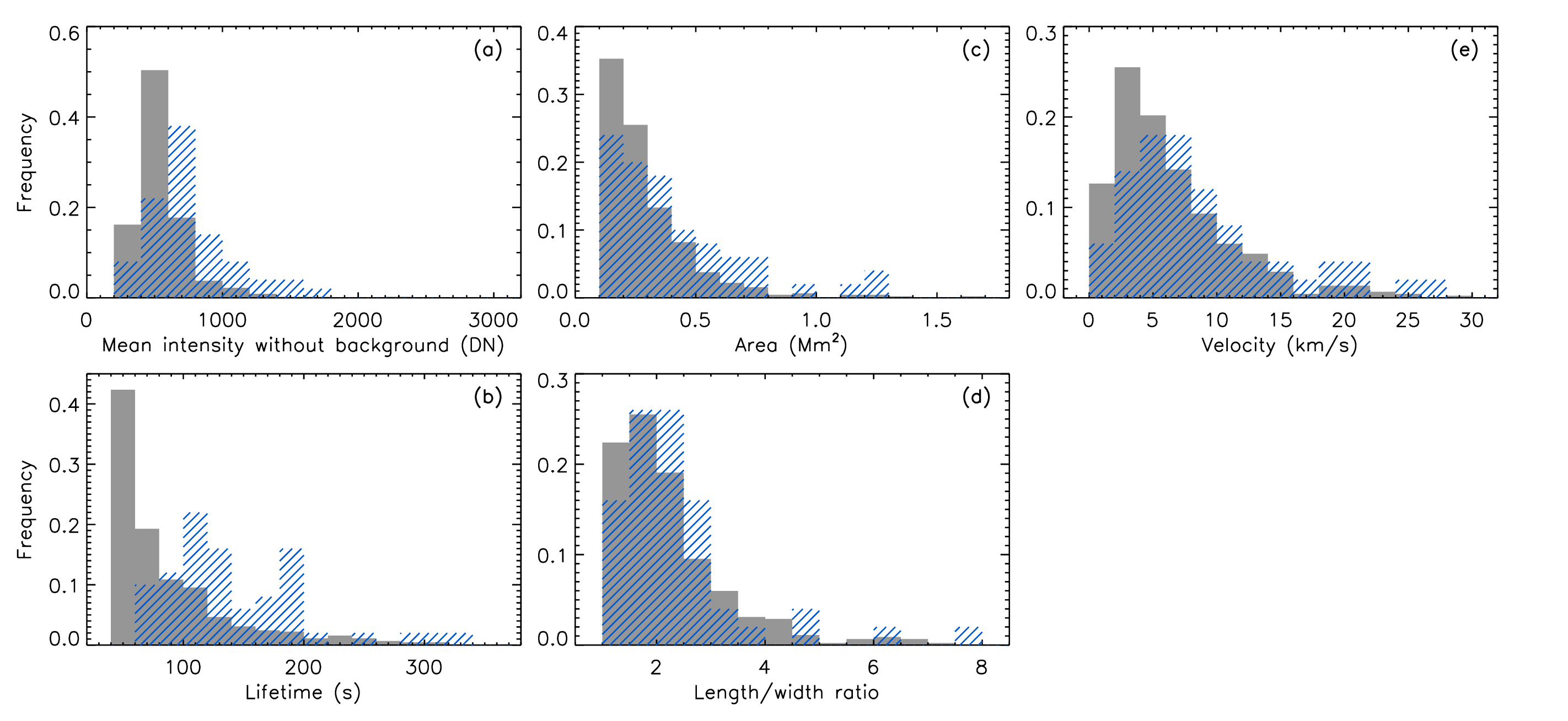}
\caption{Normalized distributions of the properties (mean intensity, lifetime, area, ratio of length to width and velocity) of the base brightenings detected with both visual identification method (50 base brightenings) and automatic method (451 base brightenings). The grey bars show the results from the automatic method and the blue line-filled bars show the results from the visual identification method.}
\label{res_fig2}
\end{figure*}

\begin{table*}[ht]
    \centering
    \caption{Automatically detected base brightenings categorized according to directions}
    \label{tab1}
    \begin{tabular}{*{4}{c}}
      \hline
         & Directional & Random-like walk & Undefined $^*$\\
      \hline
         Number of base brightenings & 38 & 222 & 191 \\
         Mean PoS velocity & 9.6\,km\,s$^{-1}$ & 6.2\,km\,s$^{-1}$ & 6.1\,km\,s$^{-1}$ \\
         Most frequent Pos velocity & 7, 13\,km\,s$^{-1}$$^{**}$ & 3\,km\,s$^{-1}$ & 3\,km\,s$^{-1}$ \\
      \hline
    \end{tabular}
    \\
    \vspace{3pt}
    \emph{$^*$} Base brightenings appearing in only 2 frames have undefined directions 
    \emph{$^{**}$} Distribution of this group shows 2 peaks
\end{table*}

Same properties are also studied for the base brightenings captured by the automatic method. In Figure~\ref{res_fig2} we show the comparison of the distributions of the properties of the base brightenings obtained by visual identification and by the automatic method. Both methods produce similar results, that is, most of the studied base brightenings are small-scale, transient and exhibit slightly elongated morphologies.

Apart from the similarities, the automatically-detected base brightenings show peaks at smaller intensity, area, lifetime and velocity. This means the automatic method is capable of picking up fainter, smaller and shorter-lived brightenings. These brightenings appear more often than the ones selected manually. Since we neglect the brightenings that only appear in one frame, the deduced lifetime cannot be shorter than 40\,s. More than 40\% of the automatically identified base brightenings possess this minimum lifetime. The distributions of length/width ratio show less difference as we do not have a preference in the shape of the base brightenings when we select them by eye, but automatically detected ones still show a slightly higher rate of circular morphology.

Most base brightenings move with a velocity of less than 10\,km\,s$^{-1}$, irrespective of the detection method. Compared with the visual identification method, the automatic method detects more base brightenings (50$\%$) moving with even lower velocity, of less than 5\,km\,s$^{-1}$. For these, the distribution peaks at 3--5\,km\,s$^{-1}$. Note that, as we only use the first and last positions of each base brightening to get an average velocity, the small velocities could be caused not only by the slow motion, but also by base brightenings moving in variable directions. We categorize all 451 base brightenings according to their directions of motion. Excluding the 191 short-lived base brightenings that appear in only two frames, for which a variation in the direction of motion cannot be defined, only 38 of the remaining ones have a relatively fixed direction of motion, i.e., the difference of their directions in two successive time intervals is always not less or equal to $\pm$ 30\textdegree. These events also exhibit relatively greater speed (see Table \ref{tab1}). Two peaks are found in their velocity distribution at 7\,km\,s$^{-1}$ and 13\,km\,s$^{-1}$. Meanwhile, the remaining 222 base brightenings show a more random-like motion.

\subsection{Base brightenings and PDs}

\begin{figure*}[tbp]
\centering 
\includegraphics[width=1.\textwidth]
{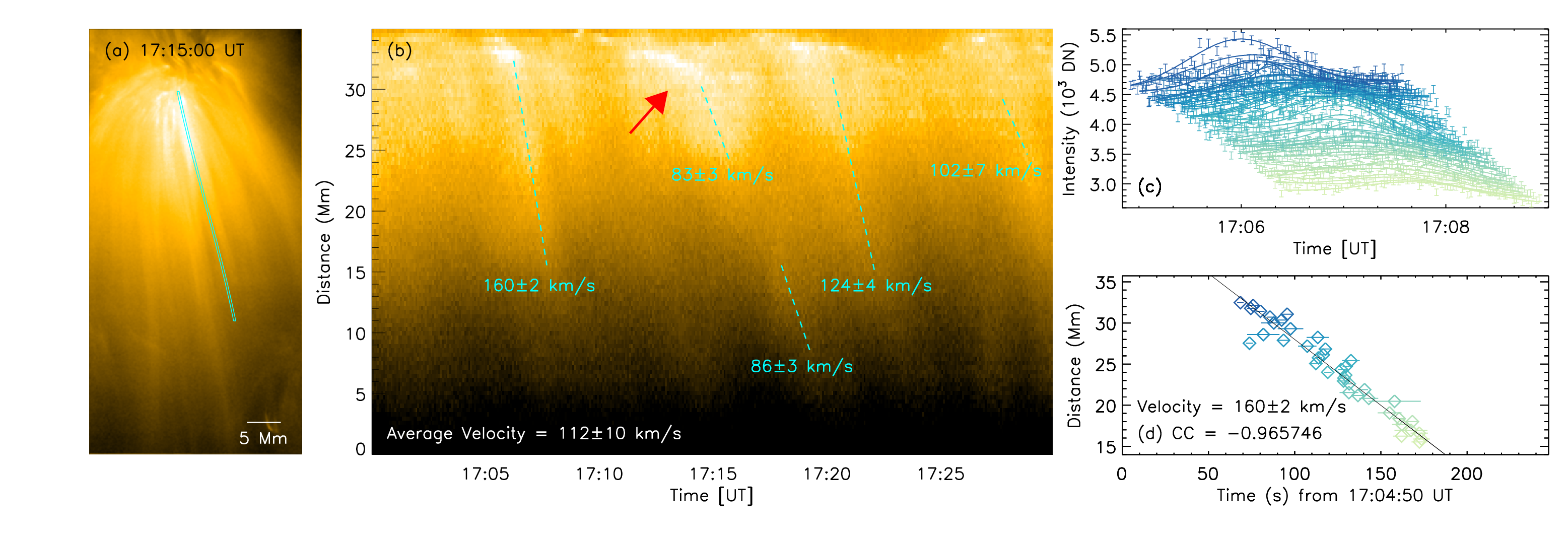}
\caption{(a) \hrieuv image of plume 2 showing the base region and plume streams. The overplotted box shows the position of the slice that is taken to form the timeslice plot shown in (b). (b) Timeslice plot. The PDs are marked with blue dashed lines. The average velocity is 112\,km\,s$^{-1}$. The red arrow points to one of the PDs, which is potentially transformed by a slow-moving base brightening. As an example, (c) shows the light curves at different distances (see different colors) of the first PD. The light curves are fitted with a Gaussian function. The central time (where the light curve peaks) derived from the fitting can be found in (d), together with the linear fitting of distances vs central time. The error of intensity in (c) is calculated taking the readout noise and dark current into account. This error is propagated to get the error of central time in (d).\\
(The associated movie is available online.)}
\label{res_fig3}
\end{figure*}

The velocities of the base brightenings are very small compared with the velocities of PDs within plumes. Previous studies suggest that the apparent outward speeds range from several tens to 300\,km\,s$^{-1}$. A similar velocity can be detected in our work as well. In Figure~\ref{res_fig3}, we take a long slit starting from the base region and following one of the bright threads in the plume to make a time-slice plot (see Figure~\ref{res_fig3}~(b)). The position of this slit is shown in Figure~\ref{res_fig3}~(a). In this time-slice plot, one can clearly see the recurring base brightenings at the footpoint and the PDs following up. We mark some of them that are easy to distinguish from the background with the straight dashed lines.

To calculate the velocities, the light curves with an original time cadence of 5\,s (and without spatial running average applied) of each PD at different distances along the slit are fitted by Gaussian functions. In Figure~\ref{res_fig3}~(c) we show the light curves of the first PD, with different colors indicating different distances. The error bars come from the readout noise and dark current (see equation S1 in \citet{chitta_picojets_2023}). In this dataset, the exposure time is 2.8\,s and the photons to DN conversion factor is 7.29\,DN\,photon$^{-1}$. The temporal positions of the PD where its light curve peaks derived from the fitting are then fitted linearly to the distance (Figure~\ref{res_fig3}~(d)). The speeds are distributed over a wide range, from 80 to 160\,km\,s$^{-1}$. The slow ones have a similar velocity as reported in \citet{gabriel_contribution_2003,gabriel_solar_2005}, while none of the PDs in our dataset have a velocity as small as the 25\,km\,s$^{-1}$ reported by \citet{fu_measurements_2014}. The LoS Doppler velocities at a height of about 20--40\,Mm have been found to be 30--60\,km\,s$^{-1}$ \citep{hassler_observations_1997,banerjee_polar_2000,teriaca_nascent_2003,wilhelm_solar_2006,banerjee_signatures_2009}. For comparison, outflow velocity of PDs and small jets as measured in \citet{pucci_birth_2014,kumar_quasi-periodic_2022, chitta_picojets_2023} is in the range of a few 10\,km\,s$^{-1}$ to a few 100\,km\,s$^{-1}$.

In one case, we observe a brightening at the plume base propagating at a speed of about 10\,km\,s$^{-1}$, potentially transitioning to a PD moving with a higher speed of about 80\,km\,s$^{-1}$ (see the arrow in Figure~\ref{res_fig3}~(b)). This could be possible evidence that the base brightenings and PDs at greater heights are intrinsically related, but one such overlap may also occur purely by chance. Also, their velocities are very different. However, as the velocities we measure here are components on the PoS, the comparison of the real velocities of the moving base brightenings and PDs should take projection effects into consideration.

\section{Velocity correction}
\label{sec-vel}

\begin{figure*}[htbp]
\centering 
\includegraphics[width=1.0\textwidth]{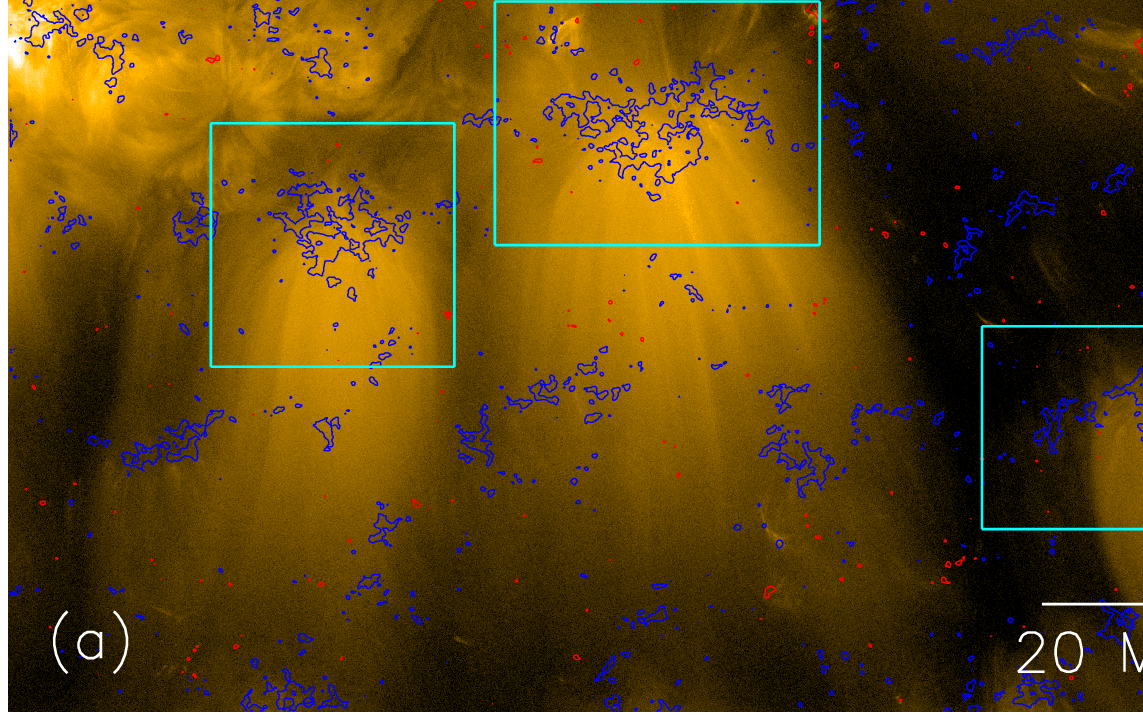}
\caption{Magnetic field imprints of the plumes. (a) Field of view covered by both \hrieuv and HRT data. The background image comes from the \hrieuv at 17:00 UT. The red (blue) contours of LoS magnetic field with the level of positive (negative) 50\,gauss are overplotted. (b), (d) and (f) show the extrapolated magnetic field lines at the location of the three plumes, the subregions marked with boxes in (a). (c), (e) and (g) are histograms of frequency of the separation angle between the local magnetic field vector and LoS of three plumes. This separation angle is obtained as average over all pixels at heights between 2000 and 4000\,km along each field line; see the yellow boxes in (b), (d) and (f).}
\label{cor_fig1}
\end{figure*}

Based on our analysis on velocities, to assess the connection between the base brightenings and PDs at higher altitudes, it is important to understand the apparent acceleration from a velocity of less than 10\,km\,s$^{-1}$ of most base brightenings to around 100\,km\,s$^{-1}$ of PDs. Since the coronal hole is adjacent to active regions, we expect that the strong magnetic field from the active regions would expand and influence the open fields, especially at a relatively high altitude. This could imply that while the open magnetic field remains nearly radial (vertical) at the lower altitudes corresponding to the height of the base brightenings, it is more horizontal at the higher altitudes guiding the PDs. The bending of magnetic field lines can cause the illusion of acceleration because only the velocity component in the PoS can be captured in images. As we mentioned before, not all base brightenings move in a fixed direction. Since the ones with random-walk-like motions are possibly related to field lines that are closed or nearly horizontal before turning outward, we here focus on the base brightenings that appear to move following the same plume threads as PDs. The aim is to de-project the observed speeds of these base brightenings to determine their true velocities and assess whether these velocities align more closely with the speeds of PDs.

An extrapolation of magnetic field can help us to understand the 3D morphology of the magnetic field and to de-project the velocity component to the real velocity, assuming that the base brightenings move along the magnetic field lines. Here we use the vector magnetic data in the solar photosphere provided by High Resolution Telescope (HRT), part of Polarimetric and Helioseismic Imager on Solar Orbiter \citep[SO/PHI;][]{solanki_polarimetric_2020}. The data is taken at 16:30 UT, 30 mins earlier than the EUI observation with a high spatial resolution of about 200\,km. In this dataset, HRT has a smaller field of view (FoV) than \hrieuv and is offset to the southeast.

After aligning \hrieuv and HRT data, the region covered in both of them is shown in Figure~\ref{cor_fig1}~(a). The background image comes from the \hrieuv at 17:00 UT and the contours of the LoS magnetic field with the level of 50\,gauss are overplotted. It is clear that the roots of all plumes are closely related to magnetic concentrations spanning about 30\,Mm. The ambiguity in the transverse field's component of the SO/PHI-HRT vector magnetogram was removed using an adapted version of the Helioseismic and Magnetic Imager \citep[HMI;][]{schou_design_2012} (on board SDO) disambiguation pipeline \citep{hoeksema_helioseismic_2014, Metcalf_Overview_2006} that implements the minimal-energy disambiguation method ME0 \citep{Metcalf_resolving_1994}. 

We calculate the radial magnetic field based on the disambiguated vector magnetic data using the transformation in \citet{gary_transformation_1990}. With the radial magnetic field as the bottom boundary, we make a potential extrapolation to the height of 256 pixels (25 Mm). Figure~\ref{cor_fig1}~(b), (d) and (f) separately show the field lines where the three plumes locate. The bottom slices are the magnetic field along Z axis, which is the local vertical direction on the Sun. The extrapolation indicates that the open field lines related to the plume grow and expand to form a funnel structure, in agreement with both observations and previous simulations. The red lines in boxes represent the direction towards Solar Orbiter.

Here, all field lines reaching the top boundary are regarded as open field lines. For each open field line, we calculate the angle between the magnetic field vector and LoS direction at different heights. In Figure~\ref{cor_fig1}(c), (e) and (g), we show the distribution of the separation angle of each field line averaged over all positions between 2000 and 4000\,km. It has been suggested by \citet{berghmans_extreme-uv_2021} that the small-scale brightenings in the quiet sun observed by \hrieuv occur at heights around 2000--4000\,km, although the brightenings at the plume base need not necessarily form at the same height as quiet sun brightenings. Depending on the morphology of each plume, the distributions peak at different angles, from 20\textdegree\ to 30\textdegree. This indicates that, the velocity can be enlarged by at most a factor of 3 in the de-projection. This is not enough to explain the difference between velocities of base brightenings and PDs. Although the fastest ones (with a velocity of 30\,km\,s$^{-1}$) could move at a speed almost reaching the high velocity of PDs, most base brightenings have a 3-dimensional velocity of less than 30\,km\,s$^{-1}$. In other words, instead of an illusion caused by perspectives, the difference between these velocities is real.

However, the same angle cannot be used for the correction of velocities detected at greater heights. Due to the limitation of HRT FoV, the adjacent active regions are not covered in this data. As mentioned before, the magnetic field from the active regions could push open field lines either towards them or away from them, depending on the magnetic polarity. This effect is missing in the potential extrapolation with HRT data. To demonstrate this effect we run another extrapolation test with data from HMI. It proves that this influence exists (Appendix \ref{app_b}), although it is hard to quantify it. In summary, we believe that the angle between the direction of the PDs and LoS is likely to be greater than that of the base brightenings. That is, part of the difference between apparent speed of PDs and base brightenings could come from the projection effect.

\section{Discussion}
\label{sec-dis}

\subsection{Link between base brightenings and PDs}

\subsubsection{Are PDs mass flows or waves?}

\begin{figure*}[htbp]
\centering 
\includegraphics[width=0.7\textwidth]{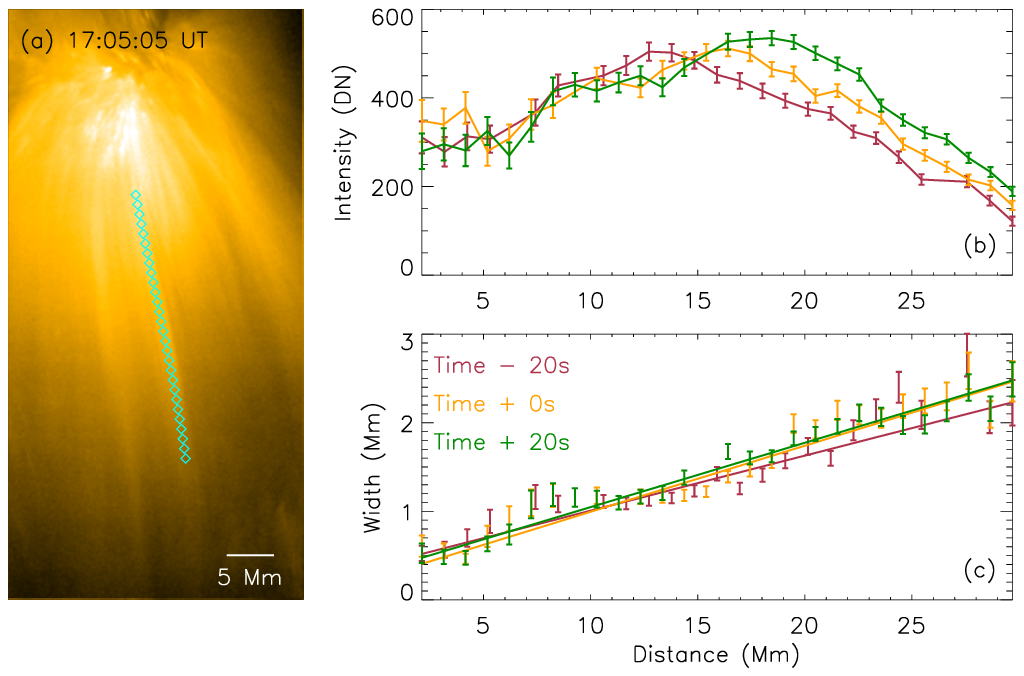}
\caption{Propagating disturbances. (a) \hrieuv image of plume 2. Positions at different distances from the base region are marked with diamonds. The intensities along the slits perpendicular to the stream direction at these positions are fitted by a Gaussian function. (b) and (c) show the peak and FWHM of the fitted Gaussian functions, as a function of the distance along the PD, at three instances (each separated by  20\,s; as identified with different colors). Same as Figure~\ref{res_fig3}~(c) and (d), the error of intensity in (b) come from the detector readout noise and dark current and the error of width is derived from the Gaussian fitting.}
\label{dis_fig1}
\end{figure*}

Let us begin the discussion by examining the physical nature of PDs, which are better understood than base brightenings and help to further investigate the potential link between the two, thereby contributing to a deeper understanding of base brightenings. In both coronal loops and plumes, PDs are widely interpreted as mass flows or waves.
Previous studies suggested the existence of slow-mode waves in plumes \citep[e.g.][]{deforest_observation_1998,krishna_prasad_omnipresent_2012}. Such slow mode waves have been invoked to be potential contributors to coronal heating \citep{deforest_observation_1998,de_moortel_observation_2000,mcewan_longitudinal_2006}. We first verify the wave interpretation using temporal variations of the intensity along plumes in our observations. To this end, we select one of the distinct PDs (see Figure~\ref{dis_fig1}~(a)) and select equally-spaced points along this PD at different distances from the base region (see diamonds in Figure~\ref{dis_fig1}~(a)). Through each point, we cut a slit perpendicular to the PD and fit a Gaussian function to the intensities along the slit. In Figure~\ref{dis_fig1}~(b), we show the amplitudes of the fitted Gaussian functions as a function of distance. A similar variation in intensity, initially increasing with height and then decaying, can be observed in other PDs as well.

This helps to characterize the nature of the PD. If the plumes are in hydrostatic equilibrium, we expect their EUV emission to fall off exponentially with certain scale-height set by the pressure at their base. In our example, we observe that the EUV emission first increases up to several megameters above the base and then decreases. Moreover, we find that the peak position of the intensity shifts outward with time (see the variations of intensity with distance at three instances separated by 20\,s). Both these factors are inconsistent with plumes being in a hydrostatic equilibrium. These factors can only be reconciled with additional heating along the plume away from the base region or excess plasma cooling from higher temperatures. But because we observe PDs propagating outward rather than downflows as typically expected in the case of plasma cooling, we suggest that our observations are consistent with a scenario of injection of material at the base of plumes. This intensity variation could also be explained by a temperature gradient along the direction of the PDs. However, This cannot be confirmed in the present study and has not been reported in previous data \citep{wilhelm_morphology_2011}.

Additionally, our analysis reveals PDs exhibiting speeds varying from 80 to 160\,km\,s$^{-1}$. Only the largest speed among them can reach the local acoustic speed $c_s \approx 150$\,km\,s$^{-1}$ at the coronal temperature $T = 10^6$\,K. Note that, because of the projection effects the apparent velocities are only a lower limit of real velocities. In an optically-thin atmosphere, it is possible that multiple PDs with similar velocities close to the local sound speed but different inclinations over the PoS could overlap along the LoS, giving the impression that different PDs have different speeds. However, considering that the slowest PDs in our sample travel at 80\,km\,s$^{-1}$, i.e., half the sound speed, it would require a separation angle from the PoS of about 60\textdegree\ to reach the typical velocity of a slow mode. Thus, the slow-mode waves could explain the propagation only when the angles between the propagating directions of different PDs are relatively large. In summary, our results are more in favor of the PDs being high speed mass flows than waves \citep[see also][]{chitta_picojets_2023}. 

\subsubsection{Mass conservation: Evolution of base brightenings into PDs}

Assuming that the PDs are mass flows, we here investigate the possibility that even the base brightenings are mass flows in their initial stage and will eventually evolve into PDs. From the trajectories and velocity distributions of the base brightenings we know that some of them, although only a small fraction (less than 10\%), have a relatively fixed moving direction with a detectable velocity, which conforms to the motion of outflows. Their velocity distribution peaks at around 7 and 13\,km\,s$^{-1}$, corresponding with the outflow velocity in coronal holes based on the Doppler shifts of transition region Ne\,{\sc viii} emission line \citep{hassler_solar_1999}. In addition, upward-propagating brightenings have also been found in AIA 304 \AA\ channel, which can correspond to the chromospheric/transition region signatures of jetlets \citep{kumar_quasi-periodic_2022,Kumar_new_2023}. The existence of elongated base brightenings also proves that some of them are jet-like structures. However, the apparent ‘outflows’ near the plume base do not share the same order of magnitude of velocity of plume flows detected at higher altitude. Despite previous works stating that ‘jetlets’ (mass flows at the plume base) could move outwards to form plumelets, in this work we find it difficult to reconcile the possibility that the base brightenings are the same evolving structures as plume outflows.

If the base brightenings evolve outwards to become plume outflows, they should have a nearly constant outward mass flow rate given by $\dot{m}=v m_\textrm{e} n_\textrm{e} A$, at different heights, where $v$ is the outflow velocity, $m_\textrm{e}$ and $n_\textrm{e}$ are the electron mass and number density, and $A$ is the cross-section area of the outflow. In the following paragraph, we estimate the changes of the velocity, the cross-section area and the number density. 

From this work, we know that we could conservatively consider an increase in velocity (from about 30\,km\,s$^{-1}$ to 100\,km\,s$^{-1}$) of a factor of 3. To estimate the expansion rate, we study the change of width of the same PD selected in Figure~\ref{dis_fig1}~(a). In Figure~\ref{dis_fig1}~(c) we show the FWHM of the fitted Gaussian functions against the distance. The expansion factor at a very low altitude is hard to be measured directly from the observation. For that the base region is too bright and dynamic to resolve plume streams. In the higher corona, where the filamentary PD substructures can be observed, we find that these streams widen linearly with distance and the stream width can increase by at least a factor of 3 at the distance of 20\,Mm relative to the base region, which gives an expansion factor of 9. To compensate the increase of the velocity and the cross-section area, a decrease in density of a factor of 27 seems to be necessary in the PD to satisfy a constant mass rate. This disagrees with our observations as the EUV emission, which depends on $n_\textrm{e}^2$ under a constant temperature assumption, does not show such a huge difference in the base region and plumelets. It is important to mention that the expansion rate can only be regarded as a rough estimation. In reality, most streams have diffused edges that can hardly be measured precisely. However, even with a negligible expansion, the base regions cannot be 3 times denser than plumes at 20--40\,Mm, which is consistent with results about plume density in the literature, as discussed below.

Different methods have been applied over the years to measure the plume density. A summary of the results of such investigations can be found in Table 3 and Figure 17 of \citet{wilhelm_morphology_2011}. The results are variable with large error bars. For instance, \citet{fludra_electron_1999} study polar plumes and find that the density decreases with height from 2\,$\times$\,10$^8$\,cm\,$^{-3}$ at the limb to 6\,$\times$\,10$^7$\,cm\,$^{-3}$ at 1.1 \rsun. \citet{young_temperature_1999} derive densities of 2.5--5.6\,$\times$\,10$^9$\,cm\,$^{-3}$ at the bright base region and of 3.8--9.5\,$\times$\,10$^8$\,cm\,$^{-3}$ above the limb at coronal temperature. Considering the power law fits, in both plume and interplume regions, $n_e$ is found to be proportional to (\rsun/R)$^8$ below R=2\rsun \citep{wilhelm_morphology_2011}. This empirically provides that the density at the plume bases is approximately 1.3 times the density at the height of 20--40\,Mm. Though different densities are found in different cases, they do not seem to meet with the possibility of mass flux conservation. 

\subsection{Decoding base brightenings}

\subsubsection{Hypothesis I: Reconnection}
One possible explanation of the base brightenings is related to magnetic reconnection. \citet{kumar_quasi-periodic_2022} selected one of the large base jets and applied a potential-field extrapolation from an HMI magnetogram to study the magnetic topology of this region. They revealed that a fan-spine structure is involved in producing the brightening and outflow. \citet{Cho_highresolutioin_2023} showed evidence of associated brightenings and blueshifted plasma (outflows) in the region with opposite magnetic polarities and found a high temperature structure at the base of the plume. \citet{Uritsky_Self_2023} identified the power-law statistical distribution of these outflows as a signature of impulsive interchange reconnection. These works support the picture of interchange reconnection between open field lines and closed loops, leading to localized rapid enhancement and ejection of materials. 

However, the necessity of this mechanism remains debated. As pointed out in the Introduction, according to MHD simulations of plumes, such an interchange reconnection scenario is not a requirement for the plume-related outflows \citep{Bora_simulation_2025}. Furthermore, previous observational studies have not discussed in detail the movement of the base brightenings. Instead, the reconnection-driven jetlets are typically observed to propagate at much higher speeds, comparable to those of the PDs analyzed in this work. It is possible that it is limited by the resolution of previous EUV imaging observations, so that the movement of these very small brightenings was not easily captured. One possibility that could account for the observed motions is that the frequent, repetitive and impulsive reconnections take place and result in discrete brightenings flashing successively over a short period of time in a small region, creating the appearance of movement. Alternatively, the movement may arise from the actual evolution of brightenings. For example, the plasma flowing along the newly-formed closed loop hits the footpoints of the loop, causing the evaporation of the chromospheric material to fill in the whole loop. In this work, we cannot corroborate whether magnetic reconnection has occurred. More magnetic field data are required to study the evolution of the photospheric magnetic field below the plumes. This hypothesis about reconnection depends on how mixed the magnetic polarities are, whose observation depends on the instrument's sensitivity and resolution.

\subsubsection{Hypothesis II: Wave-driven Type I spicules}

While different forms of reconnection, as stated above, could explain the energetic and statistical properties of the plume jetlets, the dynamic behavior of the base brightenings themselves may suggest a more complex interaction with the lower atmosphere. Specifically, the slow and more random motions of the base brightenings, distinct from the outflows suggest a potential link with direct photospheric drivers, such as spicules.

Spicules are cool and dense structures usually observed to be thin and elongated in H$\alpha$ and other chromospheric lines. \citet{de_pontieu_tale_2007} categorizes them into two types, as type I and type II spicules. Type I spicules typically have lifetimes between 3 and 7\,min and show up and downward motions along the magnetic fields with velocities of 20--50\,km\,s$^{-1}$. They are reported as the manifestations of the same driving mechanism associated with  chromospheric shocks and photospheric oscillation \citep{hansteen_dynamic_2006, de_pontieu_observations_2007}. Recently, EUV signatures of dynamic fibrils, the on-disk counterparts of Type I Spicules, have been studied using Solar Orbiter observations \citep{mandal_signatures_2023}. In contrast, type II spicules move upward rapidly with apparent speeds of 80--300\,km\,s$^{-1}$, have shorter lifetimes of 1--3\,min and do not display clear downflows \citep{de_pontieu_tale_2007,pereira_quantifying_2012,pereira_appearance_2016}

Some observational similarities can be found between base brightenings and type I spicules. Both show complex motions and have comparable velocities. Also, both of them appear to be periodic and related to chromospheric network concentrations \citep{kumar_quasi-periodic_2022}. This can help us to understand the formation of base brightenings. The type I spicules are believed to be potentially triggered by shock waves that are powered by p-modes \citep[e.g.,][]{de_pontieu_solar_2004}. This omnipresent p-mode oscillation, as also reported in \citet{kumar_quasi-periodic_2022}, can be the cause of base brightenings as well. Meanwhile, the velocities of the PDs are closer to those of type II spicules. Both of them are outwardly oriented and fade out after a certain distance. It is also reported by \citet{Cho_highresolutioin_2023} that they may have matching periodicities, suggesting a close link between them.

It is worth mentioning here that the base brightenings do not show the back-and-forth motions typical of type I spicules and dynamic fibrils. Also, the base brightening discussed here are less elongated and have a shorter lifetime than chromospheric Type I spicules. This may be a consequence of the coronal hole environment. In coronal holes, where the magnetic field is more vertical, the leakage of p-modes is more difficult due to a shorter acoustic cut-off period, forming less-elongated and shorter-lived structures, like the base brightenings observed in this study.

\section{Conclusion}
\label{sec-con}

In this study, we investigate in detail the small-scale dynamic brightenings in the base regions of three coronal hole plumes with the observation from \hrieuv. Here we find that 1) The majority of the observed brightenings at the base of plumes can be characterized by their small-scale nature (covering an area of less than 1.3\,Mm$^2$), transient behavior (with a lifespan of less than 5 minutes), and display of slightly elongated morphologies. 2) The intensities of base brightenings from different plumes are identical once the plume background is subtracted. 3) Base brightenings show complex movements. Most of them appear to move with a velocity component in the PoS of less than 10\,km\,s$^{-1}$. After de-projecting the PoS velocities to 3-dimensional velocity using an extrapolation with SO/PHI data, we find the velocities of the base brightenings are still much smaller than those of the PDs. 

For the potential explanations of the base brightenings, we first investigate the link between base brightenings and PDs, based on the two possible physical natures of the PDs, slow waves and mass flows. Assuming the constant temperature along the plume, our results are more in favor of PDs being mass flows than slow waves. In that case, our mass conservation estimation does not support the base brightenings being the initial outflows that can evolve into high-speed PDs. Given the assumptions as well as the estimation errors involved, we cannot, however, conclusively rule out the possibility that base brightenings and PDs are directly linked. But in the absence of such a connection, determining the nature of base brightenings based on an understanding of PDs remains challenging. Therefore, we propose two hypotheses on the nature of base brightenings: (i) they are associated with wave-driven Type I spicules or (ii) they are produced by interchange reconnection. Our study can provide very limited constraints on both hypotheses, highlighting the need for further detailed investigation. Considering the small size and short duration of these structures, observations with high spatial and temporal resolution are necessary and longer continuous observations could also be helpful in studying the slow-mode wave scenario in plumes. Also, the present results are derived from observations in a single bandpass and, thus, with limited temperature coverage. Future studies could take advantage of simultaneous observations from multi-instruments, i.e., photospheric magnetic fields data and spectral data, which provide information in different layers in the solar atmosphere. In this work we see that magnetic field data can assist in correcting the observed PoS velocities. In addition, Doppler velocities derived from spectral data can also provide the LoS velocity component. These are essential in understanding the mechanism of plumes and measuring the proportion of released energy that is carried by outflows to reach the upper atmosphere and eventually contribute to the solar wind.

\begin{acknowledgements}

\color{blue}
The authors thank the referee for helpful comments that have improved this paper. Solar Orbiter is a space mission of international collaboration between ESA and NASA, operated by ESA. The EUI instrument was built by CSL, IAS, MPS, MSSL/UCL, PMOD/WRC, ROB, LCF/IO with funding from the Belgian Federal Science Policy Office (BELSPO/PRODEX PEA C4000134088); the Centre National d’Etudes Spatiales (CNES); the UK Space Agency (UKSA); the Bundesministerium f\"{u}r Wirtschaft und Energie (BMWi) through the Deutsches Zentrum f\"{u}r Luft- und Raumfahrt (DLR); and the Swiss Space Office (SSO).We acknowledge funding by the Federal Ministry of Education and Research (BMBF) and the German Academic Exchange Service (DAAD).
The work of Z.H. is funded by the Federal Ministry for Economic Affairs and Climate Action (BMWK) through the German Space Agency at DLR based on a decision of the German Bundestag (Funding code: 50OU2101) and in the framework of the International Max Planck Research School (IMPRS) for Solar System Science at the Technical University of Braunschweig.
L.P.C. gratefully acknowledges funding by the European Union. Views and opinions expressed are however those of the author(s) only and do not necessarily reflect those of the European Union or the European Research Council (grant agreement No 101039844). Neither the European Union nor the granting authority can be held responsible for them. 
\color{black}

\end{acknowledgements}

%----------------------------------------------
% - use BibTeX with the regular commands:

\bibliographystyle{aa} % style aa.bst

%-------------------------------------------

\begin{appendix}

\section{Comparisons of thresholds}
\label{app_a}

In our detection methods, several thresholds have to be selected carefully. First, in both visual identification method and automatic detection, we use the DoG method, which provides the difference between the original image convoluted with two Gaussian kernels with different widths ($\sigma$). This algorithm can provide an approximation of the Laplacian of Gaussian (LoG) but is computationally more efficient. Here we use two Gaussian kernels with $\sigma=3$\,pixel and $\sigma=5$\,pixel given that a ratio around 1.6 is recommended to provide the best approximation of LoG. The intensity at each pixel of a DoG map represents the local intensity change. A DoG map may contain both positive and negative values, where positive values show bright edges or features while negative values indicate dark features or recessed edges. The zero-crossings correspond to locations of maximum gradient, that is, the edge features. In reality, zero-crossings can be very noisy and cause different brightenings to be connected, i.e., to blend with each other. Therefore, we use a reasonable threshold of 0.01 after testing, which is roughly 3-standard-deviations of the DoG image. 

\begin{figure}[htbp]
\centering 
\includegraphics[width=0.35\textwidth]{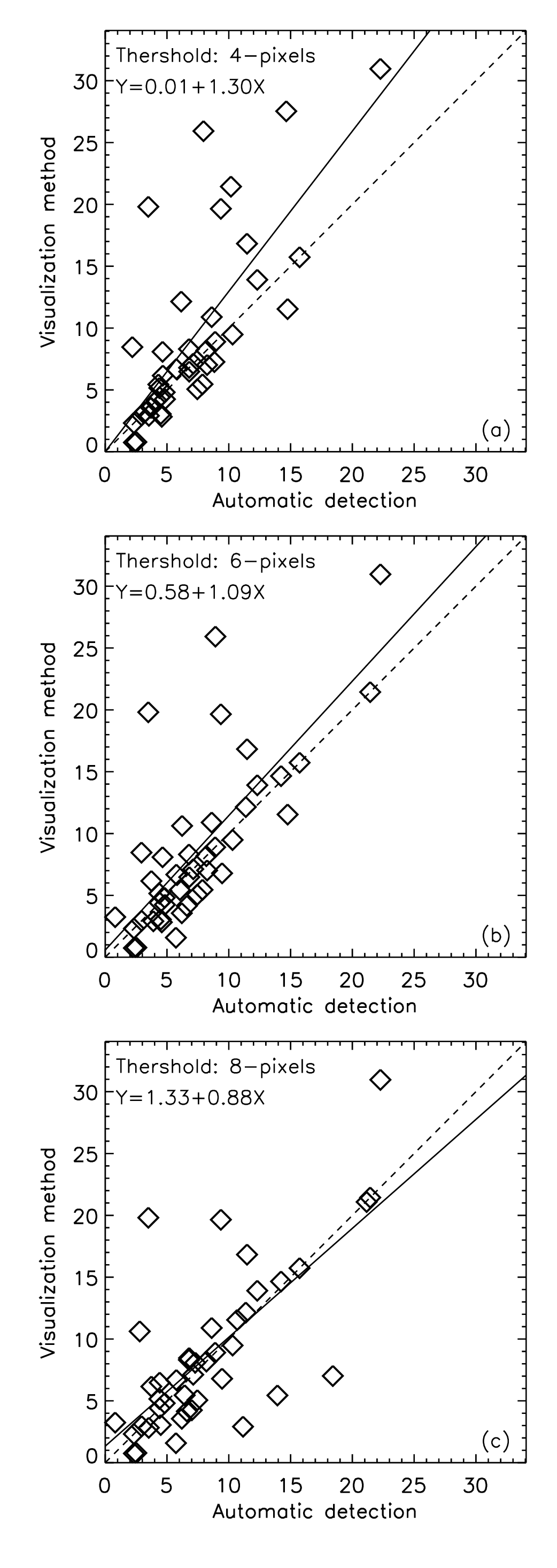}
\caption{Scatter plots showing correlations of the velocities of the same structure obtained from different methods. In (a), (b) and (c), the threshold describing the neighborhood in which we search in the following frame for the moving brightening is set to 4 pixels, 6 pixels and 8 pixels separately. The  dashed diagonals show the ideal case of equal velocities from both methods, while the solid lines are the linear fittings. The fitting parameters can be seen in the upper left corner of each panel.}
\label{app_fig1}
\end{figure}

In automatic detection, to consider evolving brightenings, we set up a threshold to define the neighborhood. In other words, in two consecutive frames, a moving structure should appear within a certain distance. The threshold setting this distance is selected based on the knowledge from visual identification method that the majority of base brightenings move with a velocity of less than 10\,km\,s$^{-1}$ but the velocity of fast ones can go up to 30\,km\,s$^{-1}$. With the pixel size of 100\,km\,pixel$^{-1}$, we test thresholds of 4, 6 and 8 pixels. As the distance threshold increases, the number of brightenings obtained decreases but the proportion of longer-lived brightenings increases. To test the reliability of different thresholds, we pick out in the automatic method the same base brightenings detected based on visual identification and compare their properties. 

\begin{figure*}[htbp]
\centering 
\includegraphics[width=1.0\textwidth]{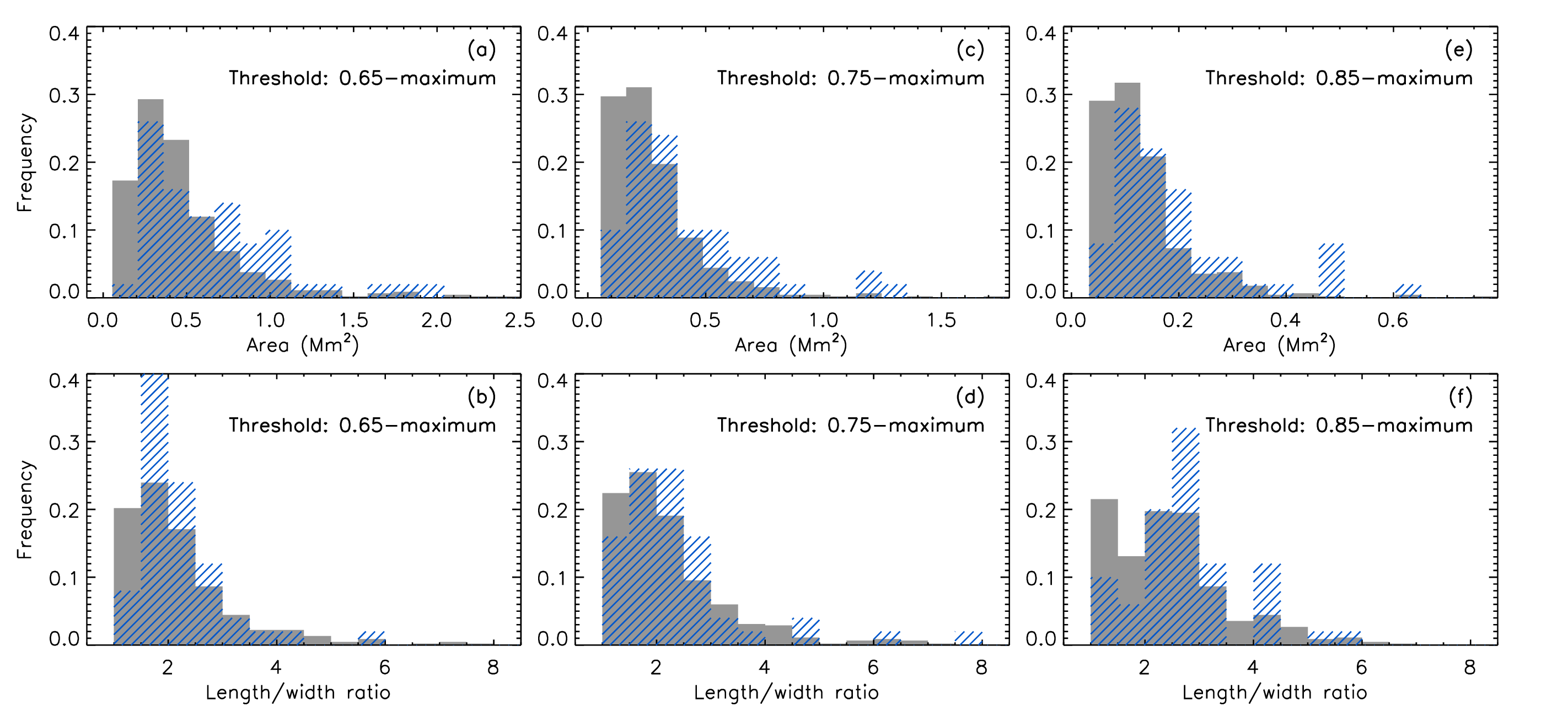}
\caption{Normalized distributions of areas (upper panels) and length/width ratio (lower panels) of the base brightenings detected in both methods. The grey bars show results from automatic method and the blue line-filled bars are results from visual identification method for comparison. From left to right panels the boundary threshold are 0.65, 0.75 and 0.85 of maximum.}
\label{app_fig2}
\end{figure*}

Not all base brightenings correspond to each other in the two methods. Differences are mainly caused by 1) two visually detected brightenings being combined by the automatic method due to spatial nearness or 2) one visually detected brightening being separated by the automatic method because of the brightening getting too faint to be marked in a middle frame. Among all the properties, the intensity matches the best and lifetime displays significant discrepancies, likely due to the short duration and the 20\,s cadence. The area and length/width ratio highly depend on the selection of the boundaries of the base brightenings. Here we show the comparison of velocities in Figure~\ref{app_fig1}. As we can see from the scatter plots, panel~(a) shows the most points deviate from $Y=X$ when the threshold is set to 4 pixels. The majority of deviations are concentrated in the upper half of the scatter plot with the threshold of both 4 pixels and 6 pixels, i.e., the velocities obtained by automatic detection are smaller compared to those from the visual identification method. This is not the case with a threshold of 8 pixels, which shows a more balanced distribution. However, by checking the image we can see that it mixes many brightenings that are spatially close. We finally chose 6 pixels as a reasonable threshold.

Another concern is about the determination of the outer boundary of each base brightening. In the first step we choose a threshold of 0.01 based on DoG maps, providing the outlines of the base brightenings. While this method works well to get the location of the base brightening, the outer boundary thus defined takes into account both the enhancement and background. This is the reason why we re-define the boundary based on the images with the local background removed. We test the thresholds of 0.65, 0.75 and 0.85 of the maximum intensity within each base brightening based on the background-removed images. Figure~\ref{app_fig2} shows the distributions of area and length/width ratio when different thresholds are selected. The area of the base brightening, although it changes significantly in value depending on the selection of the threshold, shows consistent distributions that the base brightenings with a relatively smaller area have a higher occurrence frequency. Differently, the selection of the threshold affects the distribution of the length/width ratio. With increasing threshold, more base brightenings detected by the visual identification method tend to show a relatively high length/width ratio of around 3, which does not correspond with the results from the automatic method. The distribution shows a strong peak at about 2 when the threshold equals 0.65 of the maximum. Also note that when the threshold goes down to 0.65 of maximum, the 2-dimensional Gaussian fitting has more difficulty converging, which means the shape of the base brightening is more uneven. In our work, considering all the above factors, a threshold of 0.75-maximum is used for the outlines of base brightenings. 

\section{HMI extrapolations}
\label{app_b}

We conduct another potential extrapolation test with the data from HMI to compensate for the missing of the active regions in HRT data. This experiment is done for the purpose of proving the effect of the strong magnetic field from the active regions on the directions of nearby open fields. The plumes studied in this work are located in the coronal hole close to ARs NOAA 13105 and NOAA 13107. Unfortunately, due to the separation angle between the Earth and Solar Orbiter of more than 90\textdegree, no simultaneous observation of these active regions from HMI is obtained. About one week later, the ARs and coronal hole have rotated to reach the disk center region of the sun in the FoV of HMI. However, a new AR NOAA 13127 has emerged during this time very closely to where the plumes were located. To avoid the influence of the newly emerged AR, we finally choose to use the HMI sharp$\_$cea data at 09:00 UT on 28 September 2022, which shows NOAA 13105 and NOAA 13107 in the previous rotation when they appear in the western hemisphere in HMI FoV as the Sun rotates. Despite that the ARs are stronger and the plumes in our study have not been produced at this point, this dataset could be simply used to study the effect of ARs qualitatively.

To make a comparison, we make two extrapolations. One uses the original data as the bottom boundary and another uses the processed data with the strong magnetic fields ($B_r$ above 100 Gauss) from the ARs removed. We do not change the quiet sun and coronal hole regions and some small magnetic concentrations in ARs are left for flux balance. The results can be found in Figure~\ref{app_fig3}, where the same cross-section parallel to the x-z plane is shown in both extrapolations. The streamlines show the projection of the magnetic field lines on this plane. Compared with panel (b) where there is no ARs, the streamlines on the left side of the plane in panel (a) clearly show a stronger inclination under the effect of ARs.

\begin{figure*}[htbp]
\centering 
\includegraphics[width=1.0\textwidth]{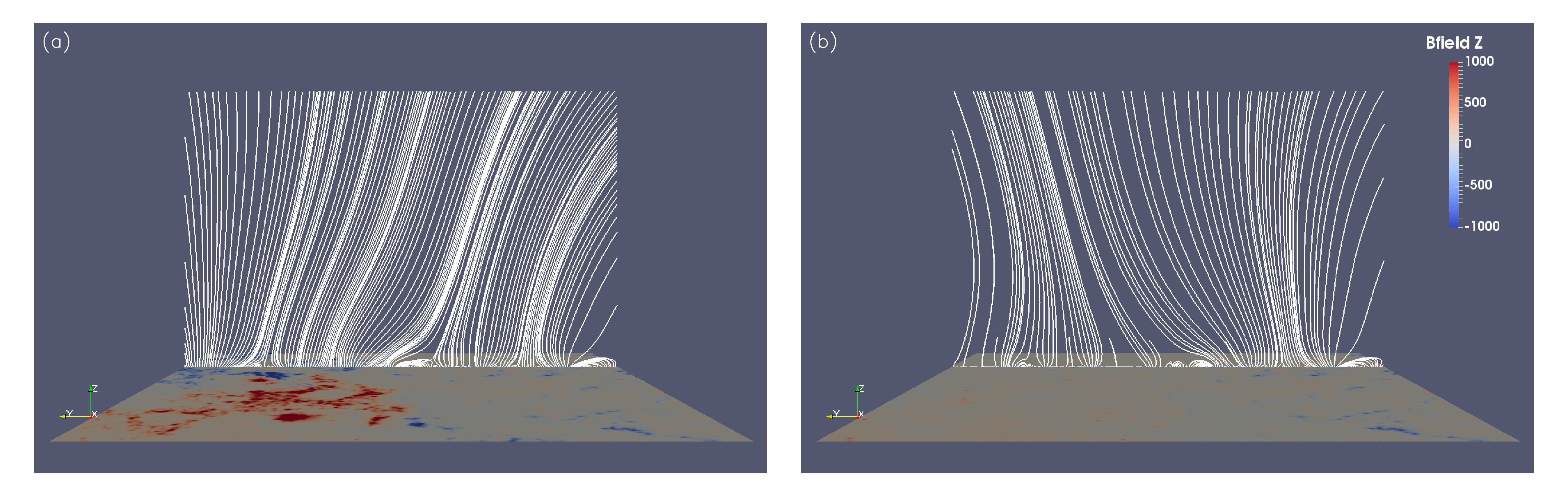}
\caption{Potential extrapolations with the original HMI data and the processed data with the strong magnetic fields from the ARs removed. One cross-section parallel to the x-z plane is selected for demonstrating the directions of magnetic fields. The streamlines show the magnetic field lines projected on this plane.}
\label{app_fig3}
\end{figure*}

\end{appendix}
 
\end{document}